\documentclass[aps, onecolumn, showpacs, amsmath]{revtex4}
\usepackage{dcolumn}
\usepackage{bm}
\usepackage{graphicx}
\usepackage{color}
\usepackage{amsthm}
\usepackage{amssymb}

\begin{document}
	
	\newcommand{\gin}[1]{{\bf\color{blue}#1}}
	\def\bc{\begin{center}}
		\def\ec{\end{center}}
	\def\bea{\begin{eqnarray}}
	\def\eea{\end{eqnarray}}
	\newcommand{\avg}[1]{\langle{#1}\rangle}
	\newcommand{\Avg}[1]{\left\langle{#1}\right\rangle}

\title{Random walks on complex networks with multiple resetting nodes: a renewal approach}
\author{Shuang Wang$^{1}$}	
\author{Hanshuang Chen$^{1}$}\email{chenhshf@ahu.edu.cn}
	
\author{Feng Huang{$^{2,3}$}}\email{huangfustc@126.com}

\affiliation{$^{1}$School of Physics and Materials Science, Anhui
		University, Hefei, 230601, China \\ $^2$Key Laboratory of Advanced Electronic Materials and Devices \& School of Mathematics and Physics, Anhui Jianzhu University, Hefei, 230601, China \\ $^3$Key Laboratory of Architectural Acoustic Environment of Anhui Higher Education Institutes, Hefei, 230601, China}

\date{\today}
	
\begin{abstract}	
Due to wide applications in diverse fields, random walks subject to stochastic resetting have attracted considerable attention in the last decade. In this paper, we study discrete-time random walks on complex network with multiple resetting nodes. Using a renewal approach, we derive exact expressions of the occupation probability of the walker in each node and mean-field first-passage time between arbitrary two nodes. All the results are relevant to the spectral properties of the transition matrix in the absence of resetting. We demonstrate our results on circular networks, stochastic block models, and Barab\'asi–Albert scale-free networks, and find the advantage of the resetting processes to multiple resetting nodes in global searching on such networks.   
		
\end{abstract}
	\maketitle

\section{Introduction}

Random walks theory on complex networks not only underlies many important stochastic dynamical processes on networked systems \cite{masuda2017random,klafter2011first}, such as epidemic spreading \cite{RevModPhys.87.925,colizza2007reaction,PhysRevX.1.011001}, population extinction \cite{WKBReview1,PhysRevLett.117.028302}, neuronal firing \cite{tuckwell1988introduction}, consensus formation \cite{PhysRevLett.94.178701}, 
but also finds a broad range of applications, such as community detection \cite{rosvall2008maps,zhou2004network,pons2005computing}, human mobility \cite{PhysRevE.86.066116,riascos2017emergence,barbosa2018human}, ranking and searching on the web \cite{PhysRevLett.92.118701,newman2005measure,lu2016vital,kleinberg2006complex,RevModPhys.87.1261}. 
In this context, two of important quantities can be identified. One is the occupation probability at stationary, which quantifies the frequency of visiting each node in the long time \cite{PhysRevLett.92.118701,PhysRevE.87.012112}.  It is well-known that the 
the stationary occupation probability is proportional to the eigenvector of transition matrix corresponding to the largest eigenvalue. For the standard random walks on connected undirected networks, the components of the eigenvector are directly proportional to degrees (or strengths for weighted networks \cite{PhysRevE.87.012112}) of nodes \cite{PhysRevLett.92.118701}. The other one is first-passage probability, that is the probability of reaching a target node for the first time. The first-passage properties underlie a wide range of stochastic processes \cite{redner2001guide,van1992stochastic,bray2013persistence}, such as fluctuation-activated transition, diffusion-limited growth, and the triggering of stock options. In particular, the mean first-passage time (MFPT) is one of quantities of interest in most of works since it gives the average lifetime of the underlying stochastic processes. In the context of complex networks, it has been shown that the MFPT on undirected networks is related to the spectral properties of the transition matrix, Besides the leading eigenmode corresponding to the largest eigenvalue that gives the stationary information, the MFPT is also related to relaxation properties of random walks \cite{PhysRevLett.92.118701,PhysRevE.87.012112,zhang2011mean,PhysRevE.79.021127,PhysRevLett.109.088701}.

Since the seminal work by Evans and Majumdar in 2011 \cite{evans2011diffusion}, random walks subject to resetting processes have received growing attention in the last decade (see \cite{evans2020stochastic} for a recent review). The walker is stochastically interrupted and reset to the initial position, and the random process is then restarted. Interestingly, the occupation probability at stationary is strongly altered. The mean  time to reach a given target for the first time can become finite and be minimized with respect to the resetting rate. Some extensions have been made in the field, such as temporally or spatially dependent resetting rate \cite{evans2011diffusion2,pal2016diffusion}, in the presence of external potential \cite{pal2015diffusion,ahmad2019first,gupta2020stochastic}, other types of Brownian motion, like run-to-tumble particles \cite{evans2018run,santra2020run,bressloff2020occupation}, active particles \cite{scacchi2018mean,kumar2020active}, and so on \cite{basu2019symmetric}.  
These nontrivial findings have triggered an enormous recent activities in the field, including statistical physics \cite{pal2017first,gupta2014fluctuating,evans2014diffusion,meylahn2015large,chechkin2018random,magoni2020ising}, stochastic thermodynamics \cite{fuchs2016stochastic,pal2017integral,gupta2020work}, chemical and biological
processes \cite{reuveni2014role,rotbart2015michaelis}, and single-particle experiments \cite{tal2020experimental,besga2020optimal}.

However, the impact of resetting on random walks in networked systems have only received  a small amount of attention \cite{avrachenkov2014personalized,avrachenkov2018hitting,rose2018spectral,PhysRevE.101.062147,christophorov2020peculiarities,PhysRevE.103.012122,lauber2021first,huang2021random}. It has been established relationships between the random walk
dynamics with resetting to one node and the spectral representation of the transition matrix in the absence of resetting \cite{rose2018spectral,PhysRevE.101.062147}. These results highlight that resetting processes is a promising way for exploring complex network topologies. 
An natural question arises: how does the resetting to multiple nodes affect the random walk dynamics on complex networks? Some related problems have been addressed in continuous one-dimensional random walks, in which the resetting position is drawn from a probability distribution, such as Gaussian distribution \cite{evans2011diffusion2,besga2020optimal}. As the ratio of the distance between the initial position and the target to the variance of the resetting position varies, an interesting dynamical phase transition occurs \cite{besga2020optimal}. When the ratio is larger than a critical value, the mean first-passage time shows a metastable optimum, whereas it disappears below the critical value. During the preparation of our manuscript, we realized that in a recent e-print \cite{arXiv:2104.00727} the authors have extended one of their previous works from a single resetting node to multiple resetting ones. By establishing the relationships between the spectrum of transition matrix in the presence of resetting and that without resetting, they derived general formula for the occupation probability at stationary and the MFPT in terms of the spectral representation of transition matrix  without resetting.

In the present work, we will use a different approach to study random walks on complex networks subject to the resetting to multiple nodes. Using a renewal approach, we derive exact expressions of the occupation probability of the walker in each node and mean-field first-passage time between arbitrary two nodes. All the results are relevant to the spectral properties of the transition matrix in the absence of resetting, that   coincides with the results of Refs.\cite{PhysRevE.101.062147,arXiv:2104.00727}. Based on the results, we apply them to circular networks, stochastic block models, and Barab\'asi–Albert (BA) scale-free networks, and find that the presence of multiple resetting nodes is more advantageous than a single resetting node in global searching on such networks.

\section{Model}\label{sec2}
We consider a particle that performing random walks on a network of size $N$ with discrete time $t=0,1,\cdots$. Supposing that at time $t$ the particle is located at node $i$, at the next time $t+1$ the particle performs either a jump to one of neighboring nodes with the probability $1-\gamma$ or a reset with the probability $\gamma$. For the former case, one of its neighboring nodes, saying node $j$, is randomly chosen with the probability $W_{ij}=A_{ij}/k_i$, where $A_{ij}$ is the element of the adjacency matrix of network, and $k_i=\sum_{i=1}^{N} A_{ij}$ is the degree of node $i$. For the latter case, the particle is reset instantaneously to a node $r_i$ with the probability $\gamma_i$ ($i=1,\cdots,R$), where $R$ the number of resetting nodes, and $\gamma=\sum_{i=1}^{R} \gamma_i$ is the total resetting probability.  
	
\section{Occupation probability}
Let us denote by $P_{ij}(t)$ the probability of finding the particle at node $j$ at time $t$ who starts from node $i$. $P_{ij}(t)$ satisfies a first renewal equation, 
\begin{eqnarray}\label{eq1}
	{P_{ij}}\left( t \right) = \Phi \left( t \right)P_{ij}^0\left( t \right) + \sum\limits_{t' = 0}^t {\sum\limits_{m = 1}^R {{\Psi _m}\left( {t'} \right){P_{{r_m}j}}\left( {t - t'} \right)} } ,
\end{eqnarray}
where $\Phi \left( t \right) = {\left( {1 - \gamma } \right)^t}$
is the probability of no reset taking place up to time $t$, and 
${\Psi _m}\left( t \right) = {\left( {1 - \gamma } \right)^{t-1}}{\gamma_m}$
is the probability of the first reset to node $r_m$ that taking place at time $t$.
$P^{0}_{ij}(t)$ is the occupation probability of the particle in the absence of resetting process (see appendix A for the derivation of $P^{0}_{ij}(t)$). The first term in Eq.(\ref{eq1}) accounts for the particle is never reset up to time $t$, and the second term in Eq.(\ref{eq1}) accounts for the particle is reset at time $t'$ for the first time, after which the process starts anew from the resetting nodes for the remaining time $t-t'$.

Let ${\kappa _{ij}}\left( t \right) = \Phi \left( t \right){P^{0}_{ij}}\left( t \right)$, and take the Laplace transform for Eq.(\ref{eq1}), $\tilde f\left( s \right) = \sum_{t = 0}^\infty  {f\left( t \right)} {e^{ - st}}$, which yields
\begin{eqnarray}\label{eq2}
	{{\tilde P}_{ij}}\left( s \right) = {{\tilde \kappa }_{ij}}\left( s \right) + \sum\limits_{m = 1}^R {{{\tilde \Psi }_m}\left( s \right){{\tilde P}_{{r_m}j}}\left( s \right)} .
\end{eqnarray}
Using the spectral decomposition for the transition matrix $W$ (see appendix A for details),  ${\tilde \kappa }_{ij}(s)$ can be written as 
\begin{eqnarray}\label{eq3}
	{\tilde \kappa _{ij}}\left( s \right) = \sum\limits_{\ell = 1}^N {\frac{{1  }}{{1 - \lambda _\ell^{}\left( {1 - \gamma } \right){e^{ - s}}}}\left\langle i \right.\left| {{\phi _\ell}} \right\rangle } \left\langle {{{\bar \phi }_\ell}} \right|\left. j \right\rangle ,
\end{eqnarray} 
where $\lambda_\ell$ is the $\ell$th eigenvalue of $\textbf{W}$, and the corresponding left eigenvector and right eigenvector are respectively $\left\langle {{{\bar \phi }_\ell}} \right|$ and $\left| {{\phi _l}} \right\rangle$, satisfying $\left\langle {{{\bar \phi }_l}} | {{\phi _m}} \right\rangle  = {\delta _{\ell m}}$ and $\sum_{\ell=1}^{N} |\phi_{\ell} \rangle \langle {\bar \phi}_{\ell}|=\textbf{I}$, where $\textbf{I}$ is an identity matrix. $\left| i \right\rangle $ denotes the canonical base with all its components equal to 0 except the $i$th one, which is equal to 1.

${{\tilde \Psi }_m}\left( s \right)$ can be expressed as
\begin{eqnarray}\label{eq4}
	{{\tilde \Psi }_m}\left( s \right) = \frac{{{\gamma _m}{e^{ - s}}}}{{1 - \left( {1 - \gamma } \right){e^{ - s}}}}.
\end{eqnarray}
	
Letting $i=r_n$ in Eq.(\ref{eq2}), we have 
\begin{eqnarray}\label{eq5}
	{{\tilde P}_{{r_n}j}}\left( s \right) = {{\tilde \kappa }_{{r_n}j}}\left( s \right) + \sum\limits_{m = 1}^R {{{\tilde \Psi }_m}\left( s \right){{\tilde P}_{{r_m}j}}\left( s \right)} .
\end{eqnarray}
Multiplying by $\tilde \Psi_n(s)$ in both sides of Eq.(\ref{eq2}) and summing over $n$ from 1 to $R$, we obtain
\begin{eqnarray}\label{eq6}
	\sum\limits_{m = 1}^R {{{\tilde \Psi }_m}\left( s \right){{\tilde P}_{{r_m}j}}\left( s \right)}  = \frac{{\sum\limits_{m = 1}^R {{{\tilde \Psi }_m}\left( s \right){{\tilde \kappa }_{{r_m}j}}\left( s \right)} }}{{1 - \sum\limits_{m = 1}^R {{{\tilde \Psi }_m}\left( s \right)} }}.
\end{eqnarray}
Subsitituting Eq.(\ref{eq6}) into Eq.(\ref{eq2}), we obtain
	\begin{eqnarray}\label{eq7}
	{{\tilde P}_{ij}}\left( s \right) = {{\tilde \kappa }_{ij}}\left( s \right) + \frac{{\sum\limits_{m = 1}^R {{{\tilde \Psi }_m}\left( s \right){{\tilde \kappa }_{{r_m}j}}\left( s \right)} }}{{1 - \sum\limits_{m = 1}^R {{{\tilde \Psi }_m}\left( s \right)} }}.
	\end{eqnarray}
Subsitituting Eq.(\ref{eq3}) and Eq.(\ref{eq4}) into Eq.(\ref{eq7}), we have 
\begin{eqnarray}\label{eq8}
	{{\tilde P}_{ij}}\left( s \right) =  \frac{{\left\langle {\left. {{{\bar \phi }_1}} \right|j} \right\rangle }}{{1 - {e^{ - s}}}} + \sum\limits_{\ell = 2}^N {\frac{{\left\langle {\left. {{{\bar \phi }_\ell}} \right|j} \right\rangle }}{{1 - {\lambda _\ell}\left( {1 - \gamma } \right){e^{ - s}}}}\left[ {\left\langle {\left. i \right|{\phi _\ell}} \right\rangle  + \frac{{{e^{ - s}}}}{{1 - {e^{ - s}}}}\sum\limits_{m = 1}^R {{\gamma _m}\left\langle {\left. {{r_m}} \right|{\phi _\ell}} \right\rangle } } \right]} .
\end{eqnarray}
Taking the inverse transform for Eq.(\ref{eq8}), we have 
\begin{equation}\label{eq9}
	{P_{ij}}\left( t \right) = \left\langle {\left. {{{\bar \phi }_1}} \right|j} \right\rangle  + \sum\limits_{\ell = 2}^N {\left[ {\left\langle {\left. i \right|{\phi _\ell}} \right\rangle \lambda _\ell^t{{\left( {1 - \gamma } \right)}^t} + \sum\limits_{m = 1}^R {{\gamma _m}\left\langle {\left. {{r_m}} \right|{\phi _\ell}} \right\rangle \frac{{1 - \lambda _\ell^t{{\left( {1 - \gamma } \right)}^t}}}{{1 - {\lambda _\ell}\left( {1 - \gamma } \right)}}} } \right]\left\langle {\left. {{{\bar \phi }_\ell}} \right|j} \right\rangle } .
\end{equation}
In stationary, $t \to \infty$, $\lambda_\ell^t \to 0$  for $\ell=2,\cdots,N$, we get to the stationary occupation probability in the presence of resetting process,
\begin{eqnarray}\label{eq10}
	{P_{j}}\left( \infty  \right) = \left\langle {\left. {{{\bar \phi }_1}} \right|j} \right\rangle  + \sum\limits_{m = 1}^R {{\gamma _m}} \sum\limits_{\ell = 2}^N {\frac{{\left\langle {\left. {{r_m}} \right|{\phi _\ell}} \right\rangle \left\langle {\left. {{{\bar \phi }_\ell}} \right|j} \right\rangle }}{{1 - {\lambda _\ell}\left( {1 - \gamma } \right)}}} ,
\end{eqnarray}
from which we see that the stationary occupation probability is independent of the starting node, but depends on the resetting nodes. The first term in Eq.(\ref{eq10}), $P_j^0\left( \infty  \right) = \left\langle {\left. {{{\bar \phi }_1}} \right|j} \right\rangle$ is stationary occupation probability in the absence of resetting process (see appendix A), and the second term in Eq.(\ref{eq10}) is an nonequilibrium contribution due to the resetting process. Eq.(\ref{eq10}) can be rewritten in the form of matrix, 
\begin{eqnarray}\label{eq11}
	{P_j}\left( \infty  \right) = \sum\limits_{m = 1}^R {{\gamma _m}} \left[ {\textbf{I} - \left( {1 - \gamma } \right)\textbf{W}} \right]_{{r_m}j}^{ - 1},
\end{eqnarray}
where $\textbf{I}$ is the $N$-dimensional identity matrix. From Eq.(\ref{eq11}), we see that ${P_j}\left( \infty  \right)$ is the weighted sum of $R$ elements of the inverse matrix of $\textbf{I}-(1-\gamma) \textbf{W}$, picked out from the $r_1, \cdots, r_R$-th rows and the $j$-th column, with the weight of each row $\gamma_m$.

\section{Mean first-passage time}
Let us suppose that there is a trap located at node $j$. Once the particle arrives at the trap, the particle is absorbed immediately. Let us denote by $Q_{ij}(t)$ the survival probability of the particle at time $t$, providing that the particle starts from node $i$. $Q_{ij}(t)$ satisfies a first renewal equation,  
\begin{eqnarray}\label{eq12}
	{Q_{ij}}\left( t \right) = \Phi \left( t \right)Q_{ij}^0\left( t \right) + \sum\limits_{t' = 0}^t {\sum\limits_{m = 1}^R {\left( {1 - {\delta _{j{r_m}}}} \right){\Psi _m}\left( {t'} \right)Q_{ij}^0\left( {t' - 1} \right){Q_{{r_m}j}}\left( {t - t'} \right)} } ,
	\end{eqnarray}
where $Q_{ij}^0\left( t \right)$ denotes the survival probability in the absence of resetting process (see appendix B for details). The fisrt term in Eq.(\ref{eq12}) corresponds to the case where there is no resetting event at all up to time $t$, which occurs with probability $\Phi(t)$. The second term in Eq.(\ref{eq12}) accounts for the event where the first resetting
to the resetting node $r_m$ that takes place at time $t'$, which occurs with probability $\Psi_m(t')$. Before the first resetting, the particle survives with probability $Q_{ij}^0\left( {t'-1} \right)$, after which the particle survives with probability ${Q_{r_m j}}\left( {t - t'} \right)$. If the resetting node is the same as the trap node, $r_m=j$, the particle is immediately absorbed as soon as it is reset. Therefore, the prefactor $1-\delta_{j{r_m}}$ ensures the second term in Eq.(\ref{eq12}) vanishes when $r_m=j$.

Let ${\chi _{ij}}\left( t \right) = \Phi \left( t \right)Q_{ij}^0\left( t \right)$, ${\eta _{ij}^{(m)}}\left( t \right) = \Psi_m \left( t \right)Q_{ij}^0\left( t-1 \right)$, (noting that $\Psi_m (0)=0$) and take the Laplace transform for Eq.(\ref{eq12}), which yields, 
\begin{eqnarray}\label{eq13}
	{{\tilde Q}_{ij}}\left( s \right) = {{\tilde \chi }_{ij}}\left( s \right) + \sum\limits_{m = 1}^R {\left( {1 - {\delta _{j{r_m}}}} \right){{\tilde \eta }^{(m)}_{ij}}\left( s \right){{\tilde Q}_{{r_m}j}}\left( s \right)} .
\end{eqnarray}
	
We take the Laplace transform for $\chi_{ij}(t)$ and $\eta_{ij}^{(m)}(t)$, which yields
\begin{eqnarray}\label{eq14}
	{{\tilde \chi }_{ij}}\left( s \right) = \sum\limits_{t = 0}^\infty  {{e^{ - st}}\Phi \left( t \right)Q_{ij}^0\left( t \right)}  = \sum\limits_{t = 0}^\infty  {{e^{ - s't}}Q_{ij}^0\left( t \right)}  = \tilde Q_{ij}^0\left( {s'} \right),
\end{eqnarray}
and 
\begin{eqnarray}\label{eq15}
	{{\tilde \eta }_{ij}^{(m)}}\left( s \right) = \sum\limits_{t = 1}^\infty  {{e^{ - st}}\Psi_m \left( t \right)Q_{ij}^0\left( t-1 \right)}  = \gamma_m e^{-s} \sum\limits_{t = 0}^\infty  {{e^{ - s't}}Q_{ij}^0\left( t \right)}  = \gamma_m e^{-s} \tilde Q_{ij}^0\left( {s'} \right),
\end{eqnarray}
where $s'=s-\ln(1-\gamma)$. Subsituting Eq.(\ref{eq14}) and Eq.(\ref{eq15}) into Eq.(\ref{eq13}), we have
\begin{eqnarray}\label{eq16}
	{{\tilde Q}_{ij}}\left( s \right) = \tilde Q_{ij}^0\left( {s'} \right) + {e^{ - s}}\tilde Q_{ij}^0\left( {s'} \right)\sum\limits_{m = 1}^R {\left( {1 - {\delta _{j{r_m}}}} \right){\gamma _m}{{\tilde Q}_{{r_m}j}}\left( s \right)} .
\end{eqnarray}
	
Letting $i=r_n$ in Eq.(\ref{eq16}) and taking the sum over $n$ from $1$ to $R$, we obtain
\begin{eqnarray}\label{eq17}
	\sum\limits_{m = 1}^R {\left( {1 - {\delta _{j{r_m}}}} \right){\gamma _m}{{\tilde Q}_{{r_m}j}}\left( s \right)}  = \frac{{\sum\limits_{m = 1}^R {\left( {1 - {\delta _{j{r_m}}}} \right){\gamma _m}\tilde Q_{{r_m}j}^0\left( {s'} \right)} }}{{1 - {e^{ - s}}\sum\limits_{m = 1}^R {\left( {1 - {\delta _{j{r_m}}}} \right){\gamma _m}\tilde Q_{{r_m}j}^0\left( {s'} \right)} }},
\end{eqnarray}
Subsituting Eq.(\ref{eq17}) into Eq.(\ref{eq16}), we obtain
\begin{eqnarray}\label{eq18}
	{{\tilde Q}_{ij}}\left( s \right) = \tilde Q_{ij}^0\left( {s'} \right) + {e^{ - s}}\tilde Q_{ij}^0\left( {s'} \right)\frac{{\sum\limits_{m = 1}^R {\left( {1 - {\delta _{j{r_m}}}} \right){\gamma _m}\tilde Q_{{r_m}j}^0\left( s' \right)} }}{{1 - {e^{ - s}}\sum\limits_{m = 1}^R {\left( {1 - {\delta _{j{r_m}}}} \right){\gamma _m}\tilde Q_{{r_m}j}^0\left( s' \right)} }}.
\end{eqnarray}
	
The MFPT from node $i$ to node $j$ is written as $\left\langle {{T_{ij}}} \right\rangle  = {{\tilde Q}_{ij}}\left( 0 \right)$, and we have by combining Eq.(\ref{eq18}) 
\begin{eqnarray}\label{eq19}
	\left\langle {{T_{ij}}} \right\rangle =\frac{{\tilde Q_{ij}^0\left( { - \ln \left( {1 - \gamma } \right)} \right)}}{{1 - \sum\limits_{m = 1}^R {\left( {1 - {\delta _{j{r_m}}}} \right){\gamma _m}\tilde Q_{{r_m}j}^0\left( { - \ln \left( {1 - \gamma } \right)} \right)} }}.
\end{eqnarray}

In terms of Eq.(\ref{eqb5}), we have 
\begin{eqnarray}\label{eq20}
	\tilde Q_{ij}^0\left( { - \ln \left( {1 - \gamma } \right)} \right) = \frac{{\sum\limits_{\ell = 2}^N {\frac{{\left\langle j \right.\left| {{\phi _\ell}} \right\rangle \left\langle {{{\bar \phi }_\ell}} \right|\left. j \right\rangle  - \left\langle i \right.\left| {{\phi _\ell}} \right\rangle \left\langle {{{\bar \phi }_\ell}} \right|\left. j \right\rangle }}{{1 - {\lambda _\ell}\left( {1 - \gamma } \right)}}}  + {\delta _{ij}}}}{{\left\langle {{{\bar \phi }_1}} \right|\left. j \right\rangle  + \gamma \sum\limits_{\ell = 2}^N {\frac{{\left\langle j \right.\left| {{\phi _\ell}} \right\rangle \left\langle {{{\bar \phi }_\ell}} \right|\left. j \right\rangle }}{{1 - {\lambda _\ell}\left( {1 - \gamma } \right)}}} }}.
\end{eqnarray}
Subsituting Eq.(\ref{eq20}) into Eq.(\ref{eq19}), we have
\begin{eqnarray}\label{eq21}
	\left\langle {{T_{ij}}} \right\rangle  = \frac{{\sum\limits_{\ell = 2}^N {\frac{{\left\langle {\left. j \right|{\phi _\ell}} \right\rangle \left\langle {\left. {{{\bar \phi }_\ell}} \right|j} \right\rangle  - \left\langle {\left. i \right|{\phi _\ell}} \right\rangle \left\langle {\left. {{{\bar \phi }_\ell}} \right|j} \right\rangle }}{{1 - {\lambda _\ell}\left( {1 - \gamma } \right)}} + {\delta _{ij}}} }}{{\left\langle {\left. {{{\bar \phi }_\ell}} \right|j} \right\rangle  + \sum\limits_{m = 1}^R {{\gamma _m}\sum\limits_{\ell = 2}^N {\frac{{\left\langle {\left. {{r_m}} \right|{\phi _\ell}} \right\rangle \left\langle {\left. {{{\bar \phi }_\ell}} \right|j} \right\rangle }}{{1 - {\lambda _\ell}\left( {1 - \gamma } \right)}}} } }}.
\end{eqnarray}
According to Eq.(\ref{eq10}), Eq.(\ref{eq21}) can be rewritten as 
\begin{eqnarray}\label{eq22}
	\left\langle {{T_{ij}}} \right\rangle  = \frac{1}{{{P_j}\left( \infty  \right)}}\left[ {\sum\limits_{\ell = 2}^N {\frac{{\left\langle {\left. j \right|{\phi _\ell}} \right\rangle \left\langle {\left. {{{\bar \phi }_\ell}} \right|j} \right\rangle  - \left\langle {\left. i \right|{\phi _\ell}} \right\rangle \left\langle {\left. {{{\bar \phi }_\ell}} \right|j} \right\rangle }}{{1 - {\lambda _\ell}\left( {1 - \gamma } \right)}} + {\delta _{ij}}} } \right].
\end{eqnarray}
In the case $j=i$, $\left\langle {{T_{ii}}} \right\rangle$ corresponds to the mean first return time to $i$, that is equal to $1/{P_i}\left( \infty  \right)$, in agreement with Kac's lemma on mean recurrence times \cite{kac1947notion}. 
	
It is also useful to quantify the ability of a process to explore
the whole network. To this purpose, we define $T(j)$ as the global MFPT (GMFPT) to the target node $j$ \cite{PhysRevE.80.065104}, averaging over the starting node $i$ with the weight equals to the stationary occupation probability, 
	\begin{eqnarray}\label{eq23}
	T\left( j \right) = \sum\limits_{i = 1}^N {{P_i}\left( \infty  \right)\left\langle {{T_{ij}}} \right\rangle } .
	\end{eqnarray}
Furthermore, one can average the GMFPT over all nodes and get a property of the whole network which was introduced as the graph MFPT (GrMFPT) \cite{PhysRevE.89.012803},
\begin{eqnarray}\label{eq23_2}
 T = \frac{1}{N}\sum_{j = 1}^N {T\left( j \right)}. 
\end{eqnarray}

\section{Applications}
\subsection{Circular networks}
We consider a circular network of size $N$, in which $\textbf{W}$ is a circulant matrix \cite{van2010graph,riascos2015fractional} with eigenvalues $\lambda_\ell=\cos[2 \pi (\ell-1)/N]$ and eigenvectors with components
	$\left\langle {\left. i \right|{\phi _l}} \right\rangle  = \frac{1}{{\sqrt N }}{e^{ - {\rm i} \left[ {2\pi \left( {\ell - 1} \right)\left( {i - 1} \right)/N} \right]}}$
	and $\left\langle {\left. {{{\bar \phi }_l}} \right|j} \right\rangle  = \frac{1}{{\sqrt N }}{e^{{\rm i}\left[ {2\pi \left( {l - 1} \right)\left( {j - 1} \right)/N} \right]}}$ (here ${\rm i}=\sqrt{-1}$) for $\ell=1,\cdots,N$. According to Eq.(\ref{eq10}), the stationary occupation probability is 
	
\begin{eqnarray}\label{eq24}
	{P_j}\left( \infty  \right) &&= \frac{1}{N} + \frac{1}{N}\sum\limits_{m = 1}^R {{\gamma _m}\sum\limits_{\ell = 2}^N {\frac{{{e^{ - i\left[ {2\pi \left( {\ell - 1} \right)\left( {{r_m} - j} \right)/N} \right]}}}}{{1 - \left( {1 - \gamma } \right)\cos \left[ {2\pi \left( {\ell - 1} \right)/N} \right]}}} } \nonumber \\ &&=\frac{1}{N} + \frac{1}{N}\sum\limits_{m = 1}^R {{\gamma _m}} \sum\limits_{\ell = 2}^N {\frac{{\cos \left( {{\varphi _\ell}{d_{{r_m}j}}} \right)}}{{1 - \left( {1 - \gamma } \right)\cos \left( {{\varphi _\ell}} \right)}}} ,
\end{eqnarray}
where ${\varphi _\ell} = 2\pi \left( {\ell - 1} \right)/N$, ${d_{{r_m}j}} = \min \left\{ {\left| {{r_m} - j} \right|,N - \left| {{r_m} - j} \right|} \right\}$ is the geodesic distance between node $r_m$ and node $j$. 
	
According to Eq.(\ref{eq22}), the MFPT can be written as
\begin{eqnarray}\label{eq25}
	\left\langle {{T_{ij}}} \right\rangle  = \frac{1}{{{P_j}\left( \infty  \right)}}\left[ {\frac{1}{N}\sum\limits_{l = 2}^N {\frac{1-{\cos \left( {{\varphi _l}{d_{ij}}} \right)}}{{1 - \left( {1 - \gamma } \right)\cos \left( {{\varphi _l}} \right)}} + {\delta _{ij}}} } \right].
\end{eqnarray}

\begin{figure}
\centerline{\includegraphics*[width=0.8\columnwidth]{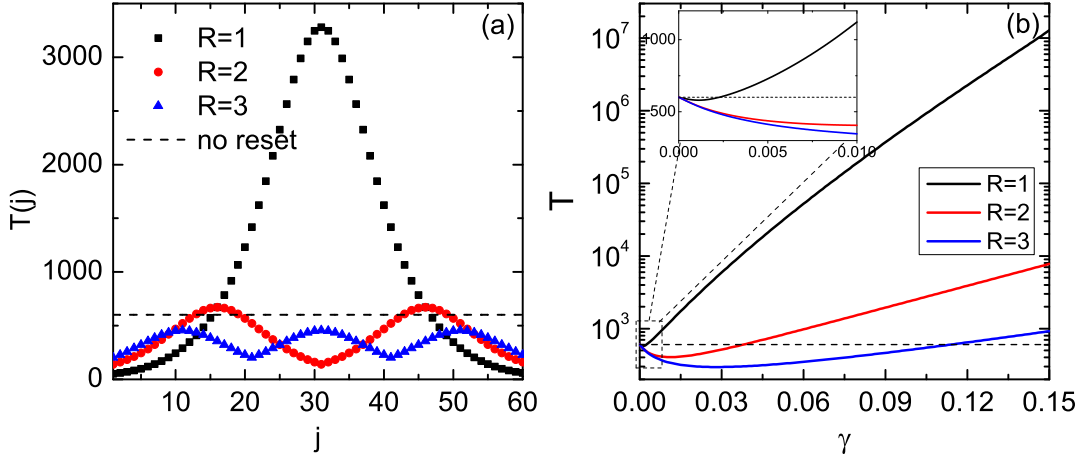}}
\caption{Results on a circular network with $N=60$ nodes. (a) The GMFPT as a function of the target node $j$ with a fixed total resetting probability $\gamma=0.01$. (b) The GrMFPT as a function of the total resetting probability $\gamma$. The inset is an enlargement in the range of $\gamma \in [0, 0.01]$. All the resetting nodes are equidistantly distributed in the network, and each resetting node has the equal resetting probability, $\gamma_m=\gamma/R$. For $R=1$, the resetting node is node 1. For $R=2$, the resetting nodes are node 1 and node 31. For $R=3$, the resetting nodes are node 1, node 21, and node 41. The horizontal dashed lines in (a) and (b) indicate the results in the absence of resetting process.}  \label{fig1}
\end{figure}

In the limit of $N \to \infty$, ${P_j}\left( \infty  \right)$ can be written as
\begin{eqnarray}\label{eq26}
	{P_j}\left( \infty  \right) = \sum\limits_{m = 1}^R {{\gamma _m}} \frac{1}{{2\pi }}\int_0^{2\pi } {\frac{{\cos \left( {{d_{{r_m}j}}\varphi } \right)}}{{1 - \left( {1 - \gamma } \right)\cos \left( \varphi  \right)}}} d\varphi .
\end{eqnarray}
Using the identity  \cite{PhysRevE.101.062147}
\begin{eqnarray}\label{id1}
	\frac{1}{{2\pi }}\int_0^{2\pi } {\frac{{\cos \left( {x\theta } \right)}}{{1 - b\cos \left( \theta  \right)}}} d\theta  = \frac{1}{{\sqrt {1 - {b^2}} }}{\left( {\frac{{1 + \sqrt {1 - {b^2}} }}{b}} \right)^{ - x}}\left( {x \ge 0,0 \le b < 1} \right), 
\end{eqnarray}
${P_j}\left( \infty  \right)$ is calculated as 
\begin{eqnarray}\label{eq27}
	{P_j}\left( \infty  \right) = \frac{1}{{\sqrt {\gamma \left( {2 - \gamma } \right)} }}\sum\limits_{m = 1}^R {{\gamma _m}} {\left[ {\frac{{1 + \sqrt {\gamma \left( {2 - \gamma } \right)} }}{{1 - \gamma }}} \right]^{ - {d_{{r_m}j}}}}.
\end{eqnarray}
In the limit of small resetting probability $0 < \gamma  \ll 1$, Eq.(\ref{eq27}) can be approximated as 
\begin{eqnarray}\label{eq28}
	{P_j}\left( \infty  \right) \approx \frac{1}{{\sqrt {2\gamma } }}\sum\limits_{m = 1}^R {{\gamma _m}} {e^{ - \sqrt {2\gamma }{d_{{r_m}j}} }} \quad {\rm for} \quad 0 < \gamma  \ll 1.
\end{eqnarray}

In the limit of $N \to \infty$, the MFPT takes the form
\begin{eqnarray}\label{eq29}
	\left\langle {{T_{ij}}} \right\rangle = \frac{1}{{{P_j}\left( \infty  \right)}}\left[ {\frac{1}{{2\pi }}\int_0^{2\pi } {\frac{{1 - \cos \left( {{d_{ij}}\varphi } \right)}}{{1 - \left( {1 - \gamma } \right)\cos \left( \varphi  \right)}}} d\varphi  + {\delta _{ij}}} \right] .
\end{eqnarray}
Using the identity $\frac{1}{{2\pi }}\int_0^{2\pi } {\frac{1}{{1 - \left( {1 - \gamma } \right)\cos \left( \varphi  \right)}}} d\varphi  = \frac{1}{{\sqrt {\gamma \left( {2 - \gamma } \right)} }}$ and combining Eq.(\ref{id1}) and Eq.(\ref{eq27}), we obtain
\begin{eqnarray}\label{eq30}
	\left\langle {{T_{ij}}} \right\rangle  = \frac{{1 - {{\left[ {\frac{{1 + \sqrt {\gamma \left( {2 - \gamma } \right)} }}{{1 - \gamma }}} \right]}^{ - {d_{ij}}}} + {\delta _{ij}}\sqrt {\gamma \left( {2 - \gamma } \right)} }}{{\sum\nolimits_{m = 1}^R {{\gamma _m}} {{\left[ {\frac{{1 + \sqrt {\gamma \left( {2 - \gamma } \right)} }}{{1 - \gamma }}} \right]}^{ - {d_{{r_m}j}}}}}}.
\end{eqnarray}
In the limit of small resetting probability $0 < \gamma  \ll 1$, Eq.(\ref{eq30}) can be approximated as 
\begin{eqnarray}\label{eq31}
	\left\langle {{T_{ij}}} \right\rangle  \approx \frac{{1 - {e^{ - \sqrt {2\gamma } {d_{ij}}}} + {\delta _{ij}}\sqrt {\gamma \left( {2 - \gamma } \right)} }}{{\sum\nolimits_{m = 1}^R {{\gamma _m}} {e^{ - \sqrt {2\gamma } {d_{{r_m}j}}}}}}.
\end{eqnarray}

In Fig.\ref{fig1}(a), we show the GMFPT as a function of the target node $j$ on a circular network with $N=60$ nodes for a fixed total resetting probability $\gamma=0.01$. For a single resetting node (node 1 is set as the resetting node), the GMFPT shows a unimodal curve with $j$. Compared with the case of no reset (see dashed line), the GMFPT can be optimized for these nodes close to the resetting node, whereas for these nodes far from the resetting node the GMFPT becomes larger. When multiple resetting nodes are set up, the GMFPT shows a multimodal variation with $j$, and the scope of optimization for the GMFPT is enlarged. For example, it is remarkable for the case of $R=3$ (the resetting nodes are node 1, node 21, and node 41), in which the GMFPT for each target node $j$ is always less than that in the case of no reset. In Fig.\ref{fig1}(b), we show the GrMFPT as a function of the total resetting probability $\gamma$. For comparison, we also show the result when the resetting process is absent, as shown by horizontal dashed line in Fig.\ref{fig1}(b). For all cases: $R=1$, $R=2$, or $R=3$, the GrMFPT always exhibits a nonmonotonic change with $\gamma$. There exists an optimal value of $\gamma$ for which the GrMFPT is minimized. For different $R$, the GrMFPT shows a quantitative difference. On the one hand, the optimal value of $\gamma$ shifts to a larger value when more resetting nodes are added. On the other hand, compared with the case of no reset, the scope of optimization for the GrMFPT (lies in below the horizontal dashed line) is expanded, embodying the advantage of multiple resetting nodes.  

\begin{figure}
	\centerline{\includegraphics*[width=0.8	\columnwidth]{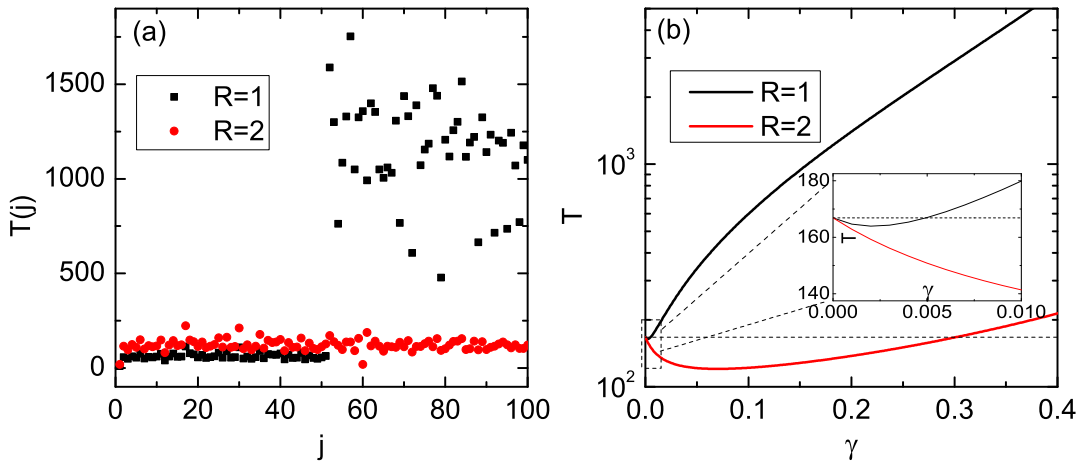}}
	\caption{Results on a stochastic block model with $N=100$ nodes with two blocks of equal sizes. (a) The GMFPT as a function of the target node $j$ with $\gamma=0.1$ (b) The GrMFPT as a function of the total resetting probability $\gamma$. The inset is an enlargement in the range of $\gamma \in [0, 0.01]$. Each resetting node has the equal resetting probability, $\gamma_m=\gamma/R$. The dashed line indicates GrMFPT in the absence of resetting.} \label{fig2}
\end{figure}

\subsection{Stochastic block model}
 We consider a stochastic block model with $N=100$ nodes, in which all nodes are divided into two blocks of equal sizes, and the connectivity probabilities within each block and inter-block are $p_{in}=0.5$ and $p_{out}=0.005$, respectively \cite{newman2018networks}. This network has obvious community structure \cite{PhysRep10000075}. Under this case, the walker tends to be trapped in a block, so that the global search becomes inefficient \cite{masuda2017random}. The difficulty may be overcome if we choose at least one resetting node in each block. Due to the resetting processes, the walker is likely to escape from one block to another. In Fig.\ref{fig2}, we show the expectation is achievable. In Fig.\ref{fig2}(a), we show the GMFPT as a function of the target node $j$ with the total resetting probability $\gamma=0.1$. For a single resetting node, $R=1$, we find that the GMFPT shows the step-like change with $j$. If the target node and the resetting node belong to the same block, the GMPTs are much less than those when they are in different blocks. If we randomly choose one resetting node in each block, $R=2$, the discrepancy in GMFPT vanishes, and the GrMFPT is thus expected to be largely reduced. In Fig.\ref{fig2}(b), we show the GrMFPT as a function of $\gamma$. The  GrMFPT exhibits a minimum at $\gamma=0.002$ for $R=1$ and at $\gamma=0.07$ for $R=2$. Comparing with the case of no resetting ($\gamma=0$), the GrMFPT can be accelerated in the range of $0<\gamma<0.005$ for $R=1$, and in the range of $0<\gamma<0.301$ for $R=2$. Therefore, the choice of multiple resetting nodes is advantageous to optimize the  GrMFPT.  
 

\subsection{BA networks}
\begin{figure}
	\centerline{\includegraphics*[width=0.6\columnwidth]{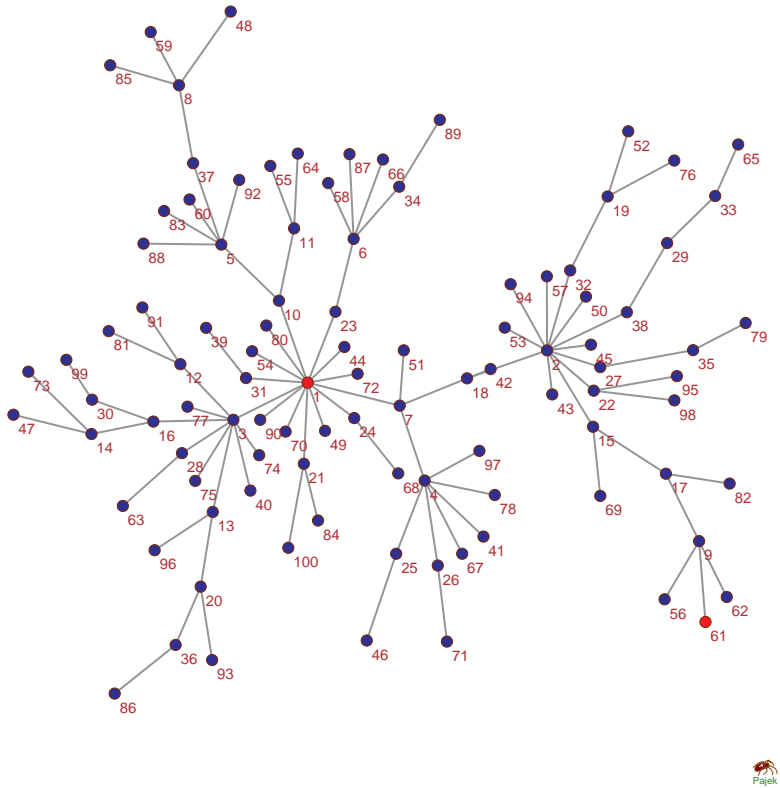}}
	\caption{An BA network with $N=100$ nodes and average degree $\left\langle k \right\rangle  = 2$, in which node 1 and node 61 are selected as the candidate resetting nodes.} \label{fig3}
\end{figure}

\begin{figure}
	\centerline{\includegraphics*[width=0.8\columnwidth]{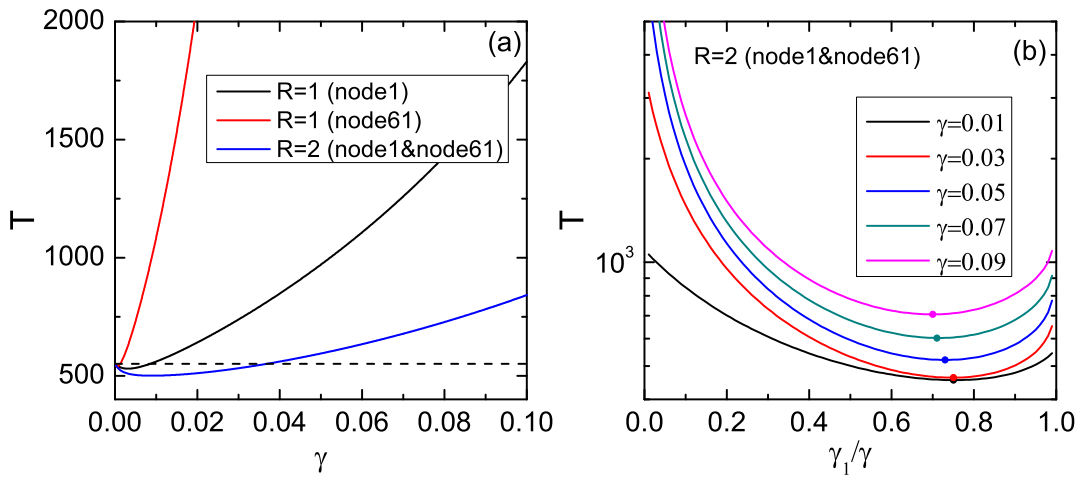}}
	\caption{Results on a BA network shown in Fig.\ref{fig3}. (a) The GrMFPT as a function of the total resetting probability $\gamma$. The dashed line indicates the result of no resetting process. For $R=2$, we have set all the resetting nodes with equal resetting probabilities, i.e., $\gamma_1=\gamma_2=\gamma/2$. (b) The GrMFPT as a function of the ratio $\gamma_1/\gamma$ for different values of $\gamma$. } \label{fig4}
\end{figure}

We consider an BA network with $N=100$ nodes and average degree $\left\langle k \right\rangle  = 2$ \cite{Science.286.509}, from which we select two nodes, node 1 and node 61, as the candidate resetting nodes, as shown in Fig.\ref{fig3}. We consider three different resetting protocols. In the first two cases, we only choose a single resetting node, corresponding to node 1 or node 61 as the only resetting node. In the three case, node 1 and node 61 are both resetting nodes. In Fig.\ref{fig4}(a), we show the GrMFPT as a function of the total resetting probability $\gamma$. For the case $R=2$, we have set the resetting probability of resetting to each resetting node to be equal, i.e., $\gamma_1=\gamma_2=\gamma/2$. For all cases, the GrMFPT exhibits non-monotonic variation with $\gamma$, with an optimal value of $\gamma$ showing up corresponding to a minimized GrMFPT. The advantage of multiple resetting nodes is clearly shown. Furthermore, for a single resetting node, choosing a node with a larger degree is more advantageous. For $R=2$, we want to optimize the GrMFPT by adjusting the relative weight of the resetting probabilities of the two resetting nodes. To the end, in Fig.\ref{fig4}(b) we show the GrMFPT as a function of the ratio $\gamma_1/\gamma$ for several different values of $\gamma$, where $\gamma_1$ is the resetting probability of node 1. It is clearly seen that there exists an optimal ratio of $\gamma_1/\gamma$ for all $\gamma$'s under consideration. The optimal $\gamma_1/\gamma$ lies in between 0.69 and 0.75 for $0.01<\gamma<0.1$, and it decreases slowly with $\gamma$.

\section{Conclusion}
In conclusion, we have studied discrete-time random walks subject to resetting processes on undirected and unweighted networks. The random walks can be interrupted and stochastically reset to either of multiple nodes with the constant probability. Using the renewal approach, we derive exact expressions of occupation probability of the walker in each node and MFPT between arbitrary two nodes. These quantities are expressed in terms of the spectral properties of transition matrix without resetting. In particular, in stationary the resetting can lead to an additional nonequilibrium contribution to the occupation probability. It is known that in the standard random walks without resetting the stationary occupation probability just depends on the leading eigenmode of transition matrix (or degrees of nodes). In the presence of resetting, the stationary occupation probability is not only relevant to full eigenmodes of transition matrix, but also depends on the choice of resetting nodes and the probabilities to reset such nodes. To explore how the efficiency of global search depends on the resetting processes, we show the GrMFPT as a function of the total resetting probability $\gamma$ on circular networks, community networks generated by stochastic block model, and BA scale-free network with degree heterogeneity. In such networks, we find that the GrMFPT shows a non-monotonic change with $\gamma$. There exists an optimal value of $\gamma$ for which the GrMFPT is minimized. Compared with the standard random walks without resetting, there is a range of $\gamma$ values for which the GrMFPT can be optimized. However, for a single resetting node, the scope of the optimization is rather narrow. Increasing the number of resetting nodes without changing the total resetting probability, such as only two resetting nodes, the GrMFPT can be significantly reduced, so that the scope of the optimization becomes wider. Therefore, we can conclude that an appropriate choice of multiple resetting nodes is beneficial to global search on networks. The present results may open up a novel way to exploring complex networks. Furthermore, we have assumed that the probability of resetting to each given resetting node is constant. However, our approach may be generalized to the case when the resetting probability is time-dependent in order to investigate the aging effect of the random walks.

\appendix
\section{Spectral decomposition for transition matrix $\textbf{W}$}
Taking the spectral decomposition for the transition matrix $\textbf{W}$, we have
\begin{eqnarray}\label{eqa1}
	\textbf{W} = \sum\limits_{\ell = 1}^N {{\lambda _\ell}\left| {{\phi _\ell}} \right\rangle } \left\langle {{{\bar \phi }_\ell}} \right|
\end{eqnarray}
where $\lambda_\ell$ is the $\ell$th eigenvalue of $\textbf{W}$, and the corresponding left eigenvector and right eigenvector are respectively $\left\langle {{{\bar \phi }_\ell}} \right|$ and $\left| {{\phi _l}} \right\rangle$, satisfying $\left\langle {{{\bar \phi }_l}} | {{\phi _m}} \right\rangle  = {\delta _{\ell m}}$ and $\sum_{\ell=1}^{N} |\phi_{\ell} \rangle \langle {\bar \phi}_{\ell}|=\textbf{I}$.

Since $\textbf{W}$ is a stochastic matrix that satisfies the sum of each row is equal to one, its maximal eigenvalue is equal to one. Without loss of generality, we let $\lambda_1=1$ and the absolute values of other eigenvalues is less than one. The right eigenvector corresponding to $\lambda_1=1$ is simply given by $\left| {{\phi _1}} \right\rangle  = {\left( {1,1, \ldots ,1} \right)^T}$. The occupation probability $P^{0}_{ij}(t)$ without resetting is given by 
\begin{eqnarray}\label{eqa2}
	{P^{0}_{ij}}\left( t \right) = \left. {\left\langle i \right.} \right|\left. {{\textbf{W}^t}} \right|\left. j \right\rangle = \sum\limits_{\ell = 1}^N {\lambda_\ell^t\left\langle i \right.\left| {{\phi _\ell}} \right\rangle } \left\langle {{{\bar \phi }_\ell}} \right|\left. j \right\rangle 
\end{eqnarray}
where $\left| i \right\rangle $ denotes the canonical base with all its components equal to 0 except the $i$th one, which is equal to 1. In the limit of $t \to \infty$, all the eigenmodes decay to zero, except for which the stationary eigenmode corresponding to $\lambda_1$. Therefore, we get to the occupation probability at stationary in the absence of resetting, ${P_j^{0}}\left( \infty  \right) = \left\langle {{{\bar \phi }_1}} \right|\left. j \right\rangle$.

For the usual random walk, $W_{ij}=A_{ij}/k_i$, and thus the transition matrix can be written as $\textbf{W}=\textbf{D}^{-1} \textbf{A}$, where $\textbf{D}={diag} \left\{ {{k_1}, \ldots ,{k_N}} \right\}$ is a diagonal matrix. $\textbf{W}$ can be rewritten as 
\begin{eqnarray}\label{eqa3}
\textbf{W}=\textbf{D}^{-1/2}\textbf{D}^{-1/2}\textbf{ A}\textbf{D}^{-1/2}\textbf{D}^{1/2}
	=\textbf{D}^{-1/2}{ \tilde{\textbf{A}} }\textbf{D}^{1/2}
\end{eqnarray}
where $\tilde{\textbf{A}}=\textbf{D}^{-1/2}\textbf{ A}\textbf{D}^{-1/2}$ is real-valued symmetric for undirected networks ($\textbf{A}=\textbf{A}^T$). Therefore, $\textbf{W}$ is diagonalizable (i.e., spectral decomposition), and the eigenvalues of $\textbf{W}$ and $\tilde{\textbf{A}}$ are the same. Letting $\left| {{\psi _\ell}} \right\rangle$ denotes the right eigenvector corresponding to the $\ell$th eigenvalue of $\tilde{\textbf{A}}$, it is not hard to verify that $\left| {{\phi _\ell}} \right\rangle  = {\textbf{D}^{ - 1/2}}\left| {{\psi _\ell}} \right\rangle$ and $\left\langle {{{\bar \phi }_\ell}} \right| = \left\langle {{\psi _\ell}} \right|{\textbf{D}^{  1/2}}$, where  
$\left\langle {{{\bar \phi }_\ell}} \right|$ and $\left| {{\phi _l}} \right\rangle$ are 
the left eigenvector and right eigenvector of the $\ell$th eigenvalue of $\textbf{W}$, respectively.

\section{Derivation of $\tilde Q_{ij}^0(s)$}
	
In the absence of resetting, the occupation probability and first passage probability satisfies the following relation, 
\begin{eqnarray}\label{eqb1}
{P^{0}_{ij}}\left( t \right) = {\delta _{t0}}{\delta _{ij}} + \sum\limits_{t' = 0}^t {{F_{ij}^{0}}\left( {t'} \right){P^{0}_{jj}}\left( {t - t'} \right)} 
\end{eqnarray}
where $F_{ij}^{0}(t)$ is the first passage probability at time $t$ in the absence of resetting process.

In the Laplace domain, we have 
\begin{eqnarray}\label{eqb2}
{{\tilde F^{0}}_{ij}}\left( s \right) = \frac{{{{\tilde P^{0}}_{ij}}\left( s \right) - {\delta _{ij}}}}{{{{\tilde P^{0}}_{jj}}\left( s \right)}}
\end{eqnarray}
In terms of Eq.(\ref{eqa2}), $P^{0}_{ij}(s)$ can be calculated as, 
\begin{eqnarray}\label{eqb3}
\tilde P_{ij}^0\left( s \right) = \sum\limits_{t = 0}^\infty  {{e^{ - st}}P_{ij}^0\left( t \right)}  = \frac{ \left\langle {{{\bar \phi }_1}} \right|\left. j \right\rangle}{{1 - {e^{ - s}}}} + \sum\limits_{\ell = 2}^N {\frac{{\left\langle i \right.\left| {{\phi _\ell}} \right\rangle \left\langle {{{\bar \phi }_\ell}} \right|\left. j \right\rangle }}{{1 - {\lambda _\ell}{e^{ - s}}}}} 
\end{eqnarray}
Since $F_{ij}^{0}(t)=Q_{ij}^{0}(t-1)-Q_{ij}^{0}(t)$ for $t \ge 1$ and $F_{ij}^{0}(0)=1-Q_{ij}^{0}(0)$ for $t =0$, we have ${{\tilde F}_{ij}^{0}}\left( s \right) =1+ \left( {{e^{ - s}} - 1} \right){{\tilde Q}_{ij}^{0}}\left( s \right)$. Therefore, we have 
\begin{eqnarray}\label{eqb4}
\tilde Q_{ij}^0\left( s \right) = \frac{{1 - \tilde F_{ij}^0\left( s \right)}}{{1 - {e^{ - s}}}} = \frac{{\tilde P_{jj}^0\left( s \right) - \tilde P_{ij}^0\left( s \right) + {\delta _{ij}}}}{{\left( {1 - {e^{ - s}}} \right)\tilde P_{jj}^0\left( s \right)}}
\end{eqnarray}
Subsituting Eq.(\ref{eqb3}) into Eq.(\ref{eqb4}), we obtain
\begin{eqnarray}\label{eqb5}
\tilde Q_{ij}^0\left( s \right) = \frac{{\sum\limits_{\ell = 2}^N {\frac{{\left\langle j \right.\left| {{\phi _\ell}} \right\rangle \left\langle {{{\bar \phi }_\ell}} \right|\left. j \right\rangle  - \left\langle i \right.\left| {{\phi _l}} \right\rangle \left\langle {{{\bar \phi }_l}} \right|\left. j \right\rangle }}{{1 - {\lambda _\ell}{e^{ - s}}}}}  + {\delta _{ij}}}}{{\left\langle {{{\bar \phi }_1}} \right|\left. j \right\rangle  + \left( {1 - {e^{ - s}}} \right)\sum\limits_{\ell = 2}^N {\frac{{\left\langle j \right.\left| {{\phi _\ell}} \right\rangle \left\langle {{{\bar \phi }_\ell}} \right|\left. j \right\rangle }}{{1 - {\lambda _\ell}{e^{ - s}}}}} }}
\end{eqnarray}
Letting $s=0$ in Eq.(\ref{eqb5}), we obtain the mean first-passage time in the absence of resetting,
\begin{eqnarray}\label{eqb6}
\left\langle {T_{ij}^0} \right\rangle  = \tilde Q_{ij}^0\left( 0 \right) = \frac{1}{{\left\langle {{{\bar \phi }_1}} \right|\left. j \right\rangle }}\left( {\sum\limits_{\ell = 2}^N {\frac{{\left\langle j \right.\left| {{\phi _\ell}} \right\rangle \left\langle {{{\bar \phi }_\ell}} \right|\left. j \right\rangle  - \left\langle i \right.\left| {{\phi _\ell}} \right\rangle \left\langle {{{\bar \phi }_\ell}} \right|\left. j \right\rangle }}{{1 - {\lambda _\ell}}}}  + {\delta _{ij}}} \right)
\end{eqnarray}

\begin{acknowledgments}
This work is supported by the National Natural Science Foundation of China (Grants No. 11875069, No 61973001) and the Key Scientific Research Fund of Anhui Provincial Education Department under (Grant No. KJ2019A0781).
\end{acknowledgments}


\end{document}